\documentclass[ preprintnumbers,amsmath,amssymb, bibnotes]{revtex4}
\usepackage{amsthm}
\usepackage{amsmath}
\usepackage{amssymb}
\usepackage{graphicx}
\usepackage{url}
\usepackage{dsfont}
\usepackage{float}
\usepackage{bm}
\usepackage{ifthen}
\usepackage[usenames,dvipsnames]{color}
\usepackage{mathrsfs}
\usepackage[colorlinks=true,citecolor=blue,urlcolor=black]{hyperref}
\usepackage{float}
\usepackage{afterpage}
\usepackage{subfigure}


\newcommand{\sket}[1]{{\ensuremath{\lvert#1\rangle}}}
\newcommand{\lket}[1]{{\ensuremath{\left\lvert#1\right\rangle}}}
\newcommand{\ket}[1]{\if@display\lket{#1}\else\sket{#1}\fi}
\newcommand{\sbra}[1]{{\ensuremath{\langle#1\rvert}}}
\newcommand{\lbra}[1]{{\ensuremath{\left\langle#1\right\rvert}}}
\newcommand{\bra}[1]{\if@display\lbra{#1}\else\sbra{#1}\fi}
\newcommand{\sbraket}[2]{{\ensuremath{\langle#1\rvert#2\rangle}}}
\newcommand{\lbraket}[2]{{\ensuremath{\left\langle#1\!\left\rvert\vphantom{#1}#2\right.\!\right\rangle}}}
\newcommand{\braket}[2]{\if@display\lbraket{#1}{#2}\else\sbraket{#1}{#2}\fi}

\newcommand{\sketbra}[2]{{\ensuremath{\lvert #1\rangle\!\langle #2\rvert}}}
\newcommand{\lketbra}[2]{{\ensuremath{\left\lvert #1\right\rangle\!\!\left\langle #2\right\rvert}}}
\newcommand{\ketbra}[2]{\if@display\lketbra{#1}{#2}\else\sketbra{#1}{#2}\fi}

\theoremstyle{plain}

\theoremstyle{definition}

\begin{document}
\title{Asymmetric de Finetti Theorem for Infinite-dimensional Quantum Systems }

\author{Murphy Yuezhen Niu\footnote{yzniu@mit.edu}}

\affiliation{Research Laboratory of Electronics, Massachusetts Institute of Technology, 77 Massachusetts Avenue, Cambridge, Massachusetts 02139, USA}
\affiliation{Department of Physics, Massachusetts Institute of Technology, Cambridge, Massachusetts 02139, USA}

\begin{abstract}
The de Finetti representation theorem for continuous variable quantum system is first developed in \cite{renner2008} to approximate an $N$-partite continuous variable quantum state  with a convex combination of independent and identical subsystems, which requires the original state to obey permutation symmetry conditioned on successful experimental verification on $ k $ of $ N $ subsystems. We generalize the  de Finetti theorem to include  asymmetric bounds on the variance of canonical observables  and biased basis selection during the verification step.  Our result thereby enables   application of  infinite-dimensional de Finetti theorem  to situations where two conjugate measurements obey different statistics, such as the security analysis of quantum key distribution protocols based on squeezed   state  against coherent attack~\cite{Niu2016}.

\end{abstract}
\maketitle

\section{Introduction}

Most information theoretic problems on large composite quantum system can be significantly simplified by introducing external structure to the system. One  widely adopted assumption is to treat a  multi-partite quantum state as a  convex combination of independent and identically distributed~(i.i.d) subsystems. This assumption cannot be justified whenever subsystems are allowed to be entangled with each other. However, in realistic quantum communication and quantum cryptography settings, a generic quantum channel  unavoidably induces various degree of entanglement between subsystems. As a result, parameters grows exponentially with system size rendering  most information theoretic tasks formidable to solve. This calls for a universal reduction from highly entangled composite system to a more tractable one.

The quantum de Finetti theorem~\cite{deFinetti1937,Renner2007} fulfills this task by  relaxing the i.i.d structure to permutation symmetry, stating that an $ N $-partite quantum state can be approximated by a convex combination of i.i.d density operators on majority of subsystems as long as the original quantum state is invariant under permutation symmetry. Unlike i.i.d structure, permutation symmetry is feasible to realize in quantum key distribution~(QKD) protocol due to the absence of absolute ordering between keys. Additionally, entanglement between subsystems is not eradicated by permutation invariance.

The discrete quantum de Finetti theorem, where subsystem dimension $ d $ has to be much smaller than $ N $, is extended by \cite{renner2008} to continuous variable quantum system with infinite subspace dimension. The difference between discrete and continuous de Finetti theorem lies in an extra experimental verification step defined by two conjugate canonical measurement operators $ \hat{X} $  and $ \hat{Y} $ on the Hilbert space $ \mathcal{H}=L^2(\mathbb{R}) $, satisfying $ [\hat{X}, \hat{Y}]=i $.
These two conjugate measurements are more naturally defined in QKD, where two parties named after Alice and Bob generate secure  keys by sharing $ n $ quantum states and performing measurements on the joint $ n $-partite quantum state. A powerful adversary~(coherent attack) is allowed to perform arbitrary joint operations on all $ n $ states during the transmission before Alice and Bob's measurements.   Among $ n $ rounds of quantum state sharing, Alice and Bob choose randomly to measure $ k $  of the $ n $-partite state with probability $ q=1/2 $ for $ \hat{X} $ and the other half for $ \hat{Y} $ for experimental verification. The   verification passes if each one of $ k $ measurement outcomes is upper bounded by a  small value. Intuitively this bound is equivalent to the requirement $ d << N $ in discrete de Finetti proof, to reduce the error probability in approximating correlations of low magnitudes with i.i.d structures without any correlations. Since many QKD protocols leverage conjugate measurements for key generation, such verification step can be easily integrated into security analysis. 

However, the asymmetry innate to the conjugate bases measurement statistics of the continuous variable quantum state has yet to be taken care of in de Finetti theorem proof. Instead of coherent state based analysis~\cite{renner2008}, if $ n $ rounds of shared quantum state are squeezed states at input,   in order to ensure high enough success probability $ \hat{X} $ and $ \hat{Y} $ should not be bounded equally in magnitude during the experimental verification. Moreover, for practical QKD realization, basis selection probability is usually not even $ q \neq 1/2 $, which could also affect the approximate i.i.d error probability. In this work, we address these two issues by allowing experimental verification to measure two conjugate basis with uneven probability and to bound the measurement outcomes with values adapted to the specific quantum state being transmitted.  We introduce squeezed state definition and squeezing operator in Sec.~\ref{Def}, and then break down the proof for the de Finetti theorem into two part in Sec.~\ref{partI} and Sec.~\ref{PartII}. We discuss the implication of our result in Sec.~\ref{summary}.

\section{Squeezed Coherent state}\label{Def}
Before proceeding with the generalized de Finetti theorem, we review the Gaussian state characterization below which is important for the proof of the theorem. Any single-mode Gaussian state can be described as a vacuum state acted upon by displacement, squeezing and rotation operators\cite{weedbrook2011gaussian} $ \ket{\alpha, \theta, r}= \hat{D}(\alpha)\hat{R}(\theta)\hat{S}(r)\ket{0}$, each  defined by
\begin{flalign}
\hat{D}(\alpha)&=\exp[\alpha \hat{a}-\alpha^{*}\hat{a}^{\dagger}]\\
\hat{R}(\theta)&=\exp[-i\theta \hat{a}^{\dagger}\hat{a}]\\
\hat{S}(r)&=\exp[r(\hat{a}^2-\hat{a}^{\dagger 2})/2]
\end{flalign}
where $ \hat{a} $ and $ \hat{a}^\dagger $ are bosonic annihilation and creation operators satisfying $ [\hat{a}, \hat{a}^\dagger]=1 $.  Without loss of generality, we set rotation angle $ \theta=0$. The quadratures defined as $ \hat{X} =\frac{\hat{a}+\hat{a}^\dagger}{2} $ and $ \hat{Y} =\frac{\hat{a}-\hat{a}^\dagger}{2I}$ are the conjugate observables to be measured in the experimental verification step of de Finetti theorem i.i.d approximation.  The variance of two quadratures for squeezed coherent state $ \ket{\alpha,  r} $ are different: 
\begin{align}\label{varianceXY}
 |\bra{} (\Delta\hat{X})^2\ket{}|^2 &= \frac{1}{4}e^{-2r}\\
  |\bra{} (\Delta\hat{Y})^2\ket{}|^2 &= \frac{1}{4}e^{2r}.
\end{align}

\begin{figure}
  \includegraphics[width=4.5cm]{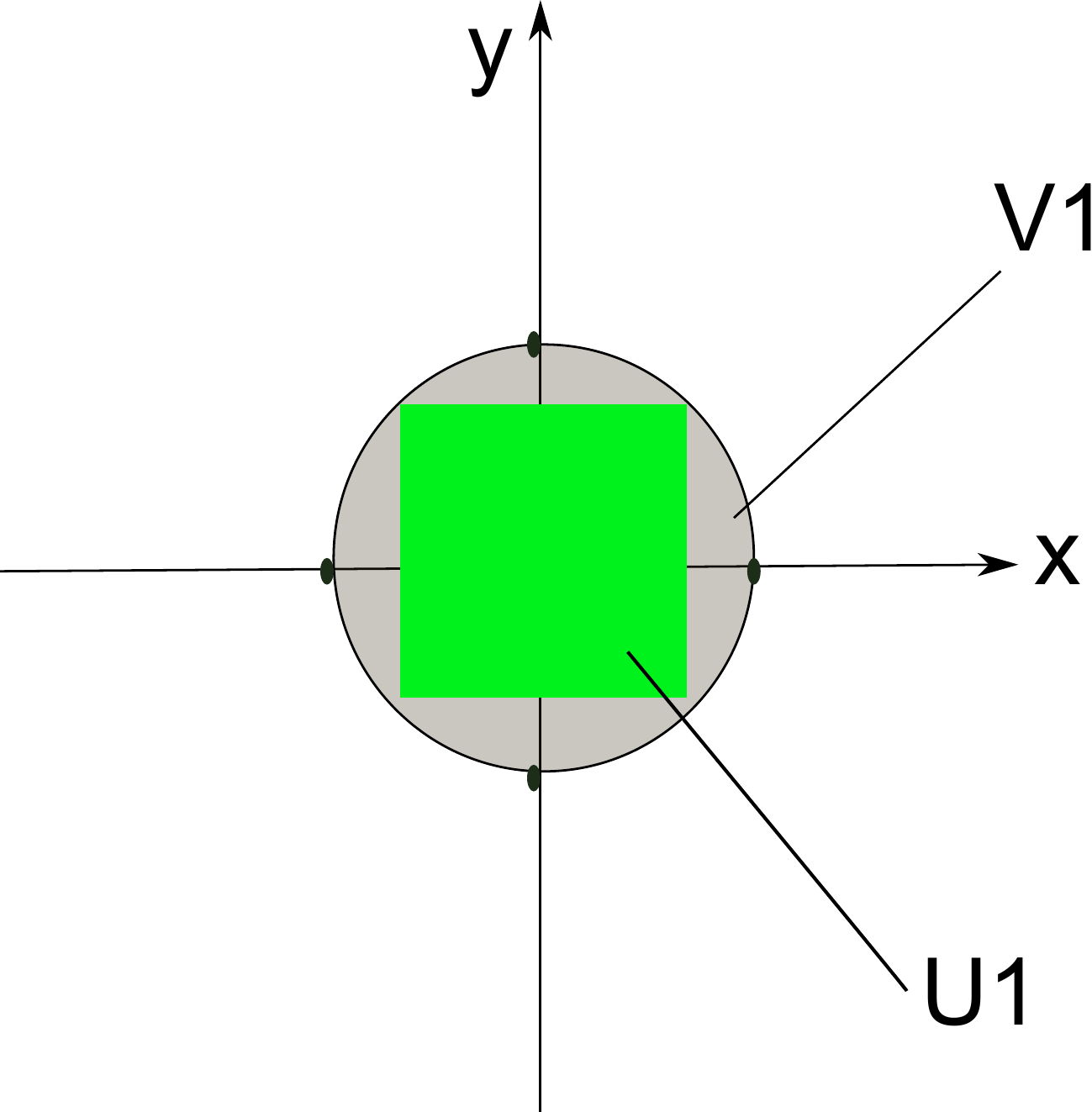}\\
  \caption{Parameter range for the measurement $ U_0:=\frac{1}{2}P^{X^2\leq n_0/2} + \frac{1}{2}P^{Y^2\leq n_0/2} $ denoted as green area,  and $V_0:= P^{\hat{X}^2 + \hat{Y}^2 \leq n_0+1}$ represented by the circle, they are complimentary to $ U_1, V_1$. The gray area measures the difference between the two.} \label{uv0}
\end{figure}

In previous analysis\cite{renner2008}, the central part of the proof  involves bounding the error probability of approximating quantum state with   $ n+k $ subsystem to one that is close to a i.i.d ensemble of $ n $ subsystems with another $ k $ subsystems on which $ \hat{X} $ or $\hat{Y}$ is chosen to be measured at random with equal probability through homodyne measurements.  The verification corresponds to the projection defined by  $ U_1:=\frac{1}{2}P^{X^2\geq n_0/2} + \frac{1}{2}P^{Y^2\geq n_0/2} $, where $ P $ denotes a projection onto the subspace of the Hilbert space $ \mathcal{\bar{H}} $ such that each quadrature measurement value of is upper bounded by a  chosen value  $ n_0 $. \cite{renner2008} proved that given the $  k$ subsystems described by the projection $ U_1 $ denoted as green square  in $ x-y $ plane in Fig.~\ref{uv0}, for  large enough $ n_0 $, the unmeasured $n$ samples can be approximated by a convex combination of i.i.d states stabilized by $ V_1:= P^{\hat{X}^2 + \hat{Y}^2 \geq n_0+1} $ bounded by the gray circle in Fig\ref{uv0} with an exponentially in $ k $ small error probability.

\begin{figure}
  \includegraphics[width=4.5cm]{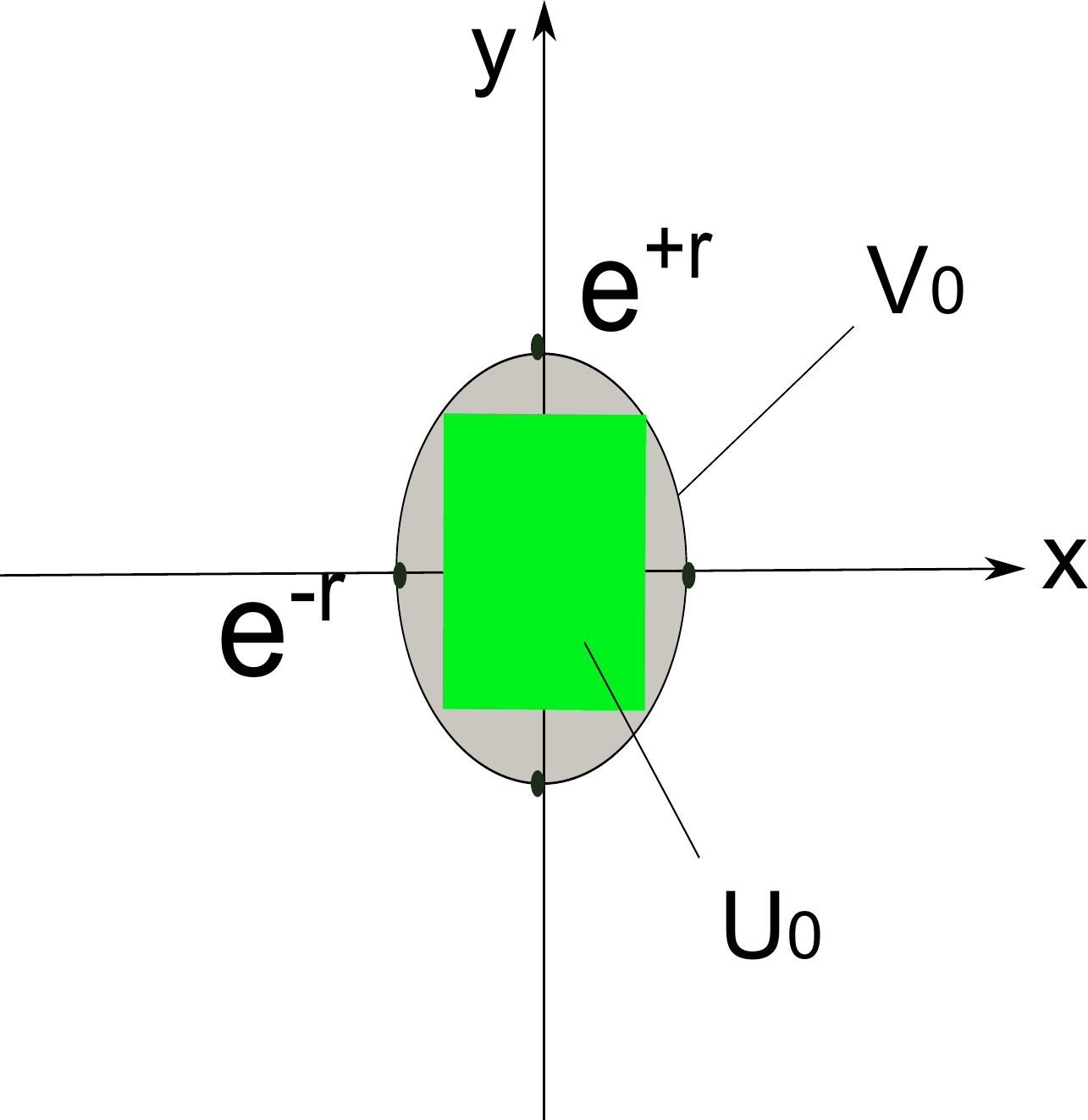}\\
  \caption{Parameter range for the measurements on squeezed coherent state $ U_0:=\frac{1}{2}P^{X^2\leq e^{-2r}n_0/2} + \frac{1}{2}P^{Y^2\leq e^{2r}n_0/2} $ denoted as green area,  and $V_0:= P^{\hat{X}^2e^{2r} + e^{-2r}\hat{Y}^2 \leq n_0+1}$ represented by the elliptical, they are complimentary to $ U_1, V_1$. The gray area measures the difference between the two.} \label{uv1}
\end{figure}

We generalize this picture to the model where two quadratures are bounded according to Eq.~(\ref{varianceXY} with the squeezing strength $ r $. And the measurements on $ \hat{X} $~(respectively $ \hat{Y} $) occurs with probability $ q $~(respectively $ 1-q $). The the   projection operator $ U_0 $  describing the state satisfied by the measurements and  the projection operator onto the inferred states $ V_0 $ are modified to:
\begin{flalign}
U_0:& =q P^{X^2\leq e^{-2r} n_0/2} + (1-q) P^{Y^2\leq e^{2r} n_0/2}\\\nonumber
V_0:& = P^{e^{2r} X^2 +  e^{-2r} Y^2 \leq n_0+1}
\end{flalign}
And the support of $ U_0$ and $V_0 $ is pictured in Fig.\ref{uv1}, which is elongated compared to \cite{renner2008}.

In order to  bound the error probability for i.i.d approximation, we need to analyze the projection operator $ U_0 $ and $ V_0 $ taking into account of an extra squeezing parameter $ r $. To complete the definition we write $ U_1= \mathrm{I}-U_0 $ and $ V_1=\mathrm{I}-V_0  $, so that $ \mathit{U}=\{ U_0, U_1\}, \mathit{V}=\{ V_0, V_1\}$ are POVMs on the Hilbert space $ \mathcal{H}$. We will prove the de Finetti theorem for squeezed state and biased basis selection in two parts below by generalizing \cite{renner2008} to squeezed state and  biased  measurement bases selection.

\subsection{de Finetti theorem- part I }\label{partI}
We define the complimentary overlap $ \gamma_{U\to V} $ as \cite{renner2008}:
\begin{eqnarray}
\gamma_{U\to V}(\delta)= \text{sup}_{\sigma}\{ \text{tr}(V\sigma): \sigma\in \mathit{S}(\mathcal{H});  \text{tr}(U \sigma)\leq \delta\}.
\end{eqnarray}
which measures the maximum probability of giving outcome V once the probability of measuring outcome U is upper bounded by $ \delta$. Denoting density operators on Hilbert space $ \mathcal{H}$  as $ \mathit{S}(\mathcal{H}) $, all permutations on $ \{1, 2,... n\} $ as $ S_n$, an arbitrary permutation as $ \pi$. Then a symmetric subspace of $ \mathcal{H}^{\otimes} $ can be constructed using a projector $ P_{Sym}^n(\mathcal{H}) $ defined as:
\begin{eqnarray}
 P_{Sym}^n(\mathcal{H})= \frac{1}{n!}\sum_{\pi\in S_n}\pi.
\end{eqnarray}
For any vector $ \mathbf{ \phi}\in Sym^n(\mathcal{H}) $ we have $ \pi\mathbf{ \phi} =\mathbf{ \phi}$. We define restricted symmetric subspace in the following for future use to connect the permutation symmetry of a subspace with a space composed of a mix of i.i.d vectors. 

Let $ P^{k+n}_{\bar{\mathcal{H}}^{\otimes n}} $ be a projector from $ \mathcal{H}^{\otimes k+n} $ on to a permutation invariant subspace $ \mathcal{H}^{\otimes k}\mathit{\bar{H}}^{\otimes n} $ composed of vectors $ \mathbf{\phi}\in \pi  \mathcal{H}^{\otimes k}\mathit{\bar{H}}^{\otimes n} $ for any given $ \pi \in S_n$, where $ \mathit{\bar{H}} $  is a subspace of the Hilbert space, and its orthogonal subspace can be denoted as $\mathit{\bar{H}}_{\perp}$. More specifically this projector can be decomposed by $ P_0=P_{\mathit{\bar{H}}} $ and $ P_1=P_{\mathit{\bar{H}}_{\perp}} $ in a permutation invariant way:
\begin{eqnarray}
 P^{k+n}_{\bar{\mathcal{H}}^{\otimes n}} = \sum_{\vec b\in\{0,1\}^{k+n}, f_{\vec b}\leq\frac{k}{k+n}}\prod_{i=1}^nP_{b_i}
\end{eqnarray}
where $\vec b $ is summed over all $ n+k  $ bit that has ones less or equal to $ k$ equivalently described by the frequency of one occurrence $ f_{\vec b}\leq\frac{k}{k+n} $. Such constraint is to guarantee that the projection on to $ \mathit{\bar{H}} ^{\otimes n}$ subspace. Since $ P^{k+n}_{\bar{\mathcal{H}}^{\otimes n}} $  is permutation invariant and commutes with any $ \pi\in S_{n+k}$, it thus also commutes with symmetric subspace projector $  P_{Sym}^{n+k} (\mathcal{H}) $. Therefore we are able to construct another symmetric projector $  P_{Sym}^{n+k} (\mathcal{H}, \bar{\mathcal{H}}^{\otimes n})= P^{k+n}_{\bar{\mathcal{H}}^{\otimes n}}P_{Sym}^{n+k} (\mathcal{H})$. This projector becomes special when the subspace $ \bar{\mathcal{H}}=\text{span}\{ \vec{\nu}\} $  so that the restricted symmetric projector onto one vector subspace writes $  Sym^{n+k} (\mathcal{H}, \vec \nu^{\otimes n})$. A density operator $ \rho^{n+k} $ belongs to the set of density operator in Hilbert space $ \mathit{S}(\mathcal{H}) $ is characterized to be \emph{almost} i.i.d if its support lies in  $  Sym^{n+k} (\mathcal{H}, \vec \nu^{\otimes n})$ for $ k\ll n$.

\noindent $ \mathbf{Lemma .1.} $ Given POVM measurements $ \mathit{U}, \mathit{V}$ on $ \mathcal{H}$, denote the $ \{ X_1,..X_{n+k} \}$ as the classical outcomes of measurement $ U^{\otimes k}\bigotimes V^{\otimes n}$. The probability that the last n bits of classical information on V measurements having more frequency of ones $ f_{X_{k+1},..., X_{k+n}} $ larger than the frequency of ones in first k bits of V measurements inferred from the first k bits of U measurements $f_{X_{1},..., X_{k}}$ is upper bounded by
\begin{widetext}
\begin{eqnarray}
P[f_{X_{k+1},..., X_{k+n}} > \gamma_{U_1 \to V_1}(f_{X_{1},..., X_{k}}+\delta)+\delta ]\leq 8 k^{3/2}e^{-k\delta^2}
\end{eqnarray}
\end{widetext}
This is proved without the specification of exact form of  $ \mathit{U}$ and $\mathit{V}$ see \cite{renner2008} and therefore will not be reproduced here. Next we are giving a bound on $ \gamma_{U_1\to V_1} $ which was initially proved using coherent state. We modify the previous proof with more general Gaussian state, i.e. squeezed coherent state $ \ket{\alpha, r}$. 

\noindent  $ \mathbf{Lemma .2.} $ Given two sets of measurements that bounds the two conjugate operators differently to fit in the squeezed state: $ U_1= q P^{\hat{X}^2< e^{-2r}n_0/2}+ (1-q) P^{\hat{Y}^2< e^{2r}n_0/2}$, and  $ V_1 =P^{e^{2r}\hat{X}^2+ \hat{Y}^2e^{-2r}< n_0} $ on the squeezed state with squeezing parameter r then $ \gamma_{U_1\to V_1}(\delta)$ is upper bounded by
\begin{widetext}
\begin{align}
\gamma_{U_1\to V_1}(\delta) & \leq\frac{2}{q(1-q)}\delta +\frac{6}{q(1-q)}[q\cdot e^r\exp(-\frac{n_0 e^{-2r}}{9})+(1-q)\cdot e^{-r}\exp(-\frac{n_0e^{2r}}{9})]
\end{align}
\end{widetext}

Proof. First, we need to show that $V_1$ is upper bounded by $W_1/2$ where $W_1$ is defined by squeezed coherent state $ \ket{\sqrt{n_0},r} $ as $ W_1 = \frac{1}{\pi} \int_{|\alpha|^2 \geq n_0} d \mu_{\alpha} \ket{\alpha, r}\bra{\alpha, r}$. The projection operator can be greatly simplified using the Bogliubov transformed number operator $ \hat{n}^{\prime}=\hat{a^{\prime}}^{\dagger} \hat{a^{\prime}}$ with $ \hat{a^{\prime}}= \cosh r \hat{a}+\sinh r\hat{a}^{\dagger} $, such that $ V_1 = \sum_{n^{\prime}=n_0}^{\infty}\ket{n^{\prime}}\bra{n^{\prime}} $ and $ W_1=\sum_n^{\prime} q_n^{\prime}\ket{n^{\prime}}\bra{n^{\prime}} $ with $  q_n^{\prime}=\Gamma(n^{\prime}+1, n_0)/\Gamma(n^{\prime}+1, 0)$, where $ \Gamma $ stands for the incomplete Gamma function~\cite{renner2008}. Using the fact that $ |\alpha|^2 \geq n_0$ and $ q_{n+1} \geq q_n  >0$, we thereby reduce  the squeezed state to coherent state with Bogliubov tranformation which gives $ V_1\leq q_{n_0}^{-1} W_1< 2W_1. $

Secondly we bound the $W_1$ with actual measurement operator $ U_1 $ which has support where the quantum state exceeds our predefined parameter region. Similar to the approach in \cite{renner2008}, we  expand our definition of  $W_1$ into a higher Hilbert space $ \mathcal{H}_1\otimes \mathcal{H}_2 $ with beam splitter operation $ \hat{B}= \exp[\frac{\pi}{4}(\hat{a_1}\otimes\hat{a_1}^{\dagger} -\hat{a_1}^{\dagger}\otimes\hat{a_1} )]$. For the first step, we show that $ \ket{f_{XY}}=\bra{0_2}\hat{S}(r)\hat{B}\ket{X}_1\otimes\ket{Y}_2 $ is the eigenstate of $ \hat{a_1}$, where $\ket{0_2} $ represents the vacuum state for mode $ \hat{a}_2 $,  $\ket{Y}_2$ is the eigenstate of operator $ \hat{Y} $ in Hilbert space $\mathcal{H}_2 $ and $ \ket{X}_1 $ denote eigenstate of $ \hat{X} $ in Hilbert space  $ \mathcal{H}_1 $:

\begin{widetext}
\begin{flalign}
\hat{a_1} \ket{f_{XY}}& =\hat{a_1} \bra{0_2}\hat{S}(r)\hat{B}\ket{X_1}\otimes\ket{Y_2} \\
&= \bra{0_2}(\hat{a_1}+\hat{a_2}^{\dagger})\hat{S}(r)\hat{B}\ket{X_1}\otimes\ket{Y_2}\\
&= \bra{0_2}\hat{S}(r)\hat{S}^{\dagger}(r)(\hat{a_1}+\hat{a_2}^{\dagger})\hat{S}(r)\hat{B}\ket{X_1}\otimes\ket{Y_2}\\
&= \bra{0_2}\hat{S}(r)[\cosh r (\hat{a_1}+\hat{a_2}^{\dagger})- \sinh r (\hat{a_1}^{\dagger}+ \hat{a_2})]\hat{B}\ket{X_1}\otimes\ket{Y_2}\\
&= (e^{-r}X+ ie^r Y) \ket{f_{XY}},
\end{flalign}
 \end{widetext}

The last equation is true from the relation $ (\hat{a_1}+\hat{a_2})\hat{B}\ket{X_1}\otimes\ket{Y_2}= \hat{B}(X_1 + iY_2)\ket{X_1}\otimes\ket{Y_2}$. Therefore $ \ket{f_{XY}} $ is a squeezed coherent state with squeezing parameter r. Following which,  $W_1$ can be redefined as
\begin{flalign}
W_1&= \int dX dY \ket{ f_{XY}}\bra{ f_{XY}}\\\nonumber
&= \int  dX dY \bra{0_2}\hat{S}(r)\hat{B}\ket{X_1}\bra{X_1}\otimes\ket{Y_2}\bra{Y_2}\hat{B}^{\dagger}\hat{S}^{\dagger}(r)\ket{0_2}
\end{flalign}
with integration range $ X^2+Y^2\geq n_0$. We can upper bound the $ W_1 $ with oppositely squeezed operators $ A $ and $ C $ defined as with squeezing parameters $ r $ and $ -r $ 
\begin{flalign}
&A=\int dX dX^{\prime} \bra{0^{\prime}}S(r) B \ket{X}\bra{X}\otimes\ket{X^{\prime}}\bra{X^{\prime}}\hat{B}^{\dagger}\hat{S}^{\dagger}(r)\ket{0^{\prime}}\\\nonumber
&C=\int dY dY^{\prime} \bra{0^{\prime}}S(-r) B \ket{Y}\bra{Y}\otimes\ket{Y^{\prime}}\bra{Y^{\prime}}\hat{B}^{\dagger}\hat{S}^{\dagger}(-r)\ket{0^{\prime}}\\\nonumber
\end{flalign}
where the integration range is $ |X^2|<n_0/2 $ and $ -\infty< X^{\prime} <\infty$, $ |Y^2|<n_0/2 $ and $ -\infty< Y^{\prime} <\infty$. Notice that the squeezing is absorbed by the squeezing operator and does not affect the integration range. We also know that $  B \ket{T}\otimes \ket{X^{\prime}} =\ket{(X+X^{\prime})\sqrt{2}}\otimes \ket{(X -X^{\prime})\sqrt{2}}$, and therefore apply the change of variable as $ x_-=(X-X^{\prime})\sqrt{2} $ and $ x=\sqrt{2}X$, the operator A can be rewritten as
\begin{flalign}
A&=\int dX |\bra{0^{\prime}}S(r) \ket{(X+X^{\prime})\sqrt{2}}|^2\int_{-\infty}^{\infty} \ket{x_-}\bra{x_-}d x_-\\\nonumber
&=\frac{1}{\sqrt{\pi}}\int_{|x|^2\geq n_0 }dt \exp[ e^{-2r}(x- x_-)^2]\int_{-\infty}^{\infty} \ket{x_-}\bra{x_-}d x_-\\\nonumber
&= F(x_-)
\end{flalign}
The second equation uses the fact that $ S^{\dagger}(r) \hat{X}S(r)= e^r \hat{X} $ and the equality $ |\bra{0}\ket{X}|^2= \exp(-X^2)/\sqrt{\pi}$. Similarly, we define $G(y_-)$ as
\begin{flalign}
C&=\int dY |\bra{0^{\prime}}S^{\dagger}(r) \ket{(Y+Y^{\prime})\sqrt{2}}|^2\int_{-\infty}^{\infty} \ket{y_-}\bra{y_-}d y_-\\\nonumber
&=\frac{1}{\sqrt{\pi}}\int_{|y|^2\geq n_0 }dt \exp[ e^{2r}(y- y_-)^2]\int_{-\infty}^{\infty} \ket{y_-}\bra{y_-}d y_-\\\nonumber
&= G(y_-)
\end{flalign}
Subsequently, $ F(x_-) $ and $ G(y_-) $ operators are bounded by 
\begin{widetext}
\begin{flalign}
&F(T)\leq P^{X^2\geq e^{-2r}a}+ F(a)< P^{X^2\geq e^{-2r}a^2}+\frac{1}{\sqrt{\pi}}\frac{\exp[-e^{-2r}(\sqrt{n_0}-a)^2]}{e^{-r}(\sqrt{n_0}-a)}\\
&G(W)\leq P^{Y^2\geq e^{2r}a}+ F(a)< P^{Y^2\geq e^{2r}a^2}+\frac{1}{\sqrt{\pi}}\frac{\exp[-e^{2r}(\sqrt{n_0}-a)^2]}{e^{r}(\sqrt{n_0}-a)}\\
\end{flalign}
\end{widetext}
where we use the fact that $ F(a)<\frac{1}{\sqrt{\pi}}\frac{\exp[-e^{-2r}(\sqrt{n_0}-a)^2]}{e^{-r}(\sqrt{n_0}-a)}$ and $ G(a)<\frac{1}{\sqrt{\pi}}\frac{\exp[-e^{2r}(\sqrt{n_0}-a)^2]}{e^{r}(\sqrt{n_0}-a)} $ with $ a\in [0, \sqrt{n_0}] $. Knowing that $ S^{\dagger}(r)=S(-r)$, we have the inequality below given binary probability distribution $ \{q, 1-q\} $ with $ q\in \{ \mathsf{R}| q\leq 1  \}$ and $ a=\sqrt{n_0/2} $: 
\begin{widetext}
\begin{flalign}
&\sqrt{q(1-q)}W_1 \leq q \cdot A + (1-q)\cdot C  \\\nonumber
&q \cdot A + (1-q)\cdot C < q\cdot P^{X^2\geq e^{-2r}n_0^2/2} + (1-q)P^{Y^2\geq e^{2r}n_0^2/2} +\frac{3}{\sqrt{\pi}\sqrt{n_0}}[q\cdot e^r\exp(-\frac{n_0e^{-2r}}{9})+(1-q)\cdot e^{-r}\exp(-\frac{n_0e^{2r}}{9})]\\\nonumber
&\rightarrow  V_1 <\frac{2}{q(1-q)}U_1 +\frac{6}{q(1-q)}[q\cdot e^r\exp(-\frac{n_0 e^{-2r}}{9})+(1-q)\cdot e^{-r}\exp(-\frac{n_0e^{2r}}{9})]
\end{flalign}
\end{widetext}
Therefore the gamma function is bounded by 
\begin{widetext}
\begin{eqnarray}
\gamma_{U_1\to V_1}<\frac{2}{q(1-q)}\delta +\frac{6}{q(1-q)}[q\cdot e^r\exp(-\frac{n_0 e^{-2r}}{9})+(1-q)\cdot e^{-r}\exp(-\frac{n_0e^{2r}}{9})]
\end{eqnarray}
\end{widetext}
which differs from the continuous variable using symmetric quadrature measurement by the extra squeezing parameter r and biased probability in bases selection $ 0<q<1$. Utilizing this new bound we complete the proof of de Finetti theorem in the next subsection.
\subsection{de Finetti theorem- part II }\label{PartII}

\noindent $ \mathbf{Lemma .3.}$ Given two conjugate operators $ \hat{X} $ and $ \hat{Y}$ on $ \mathcal{H}$, define the subspace $\tilde{\mathcal{H} } $ of the Hilbert space $\mathcal{H}  $ as the support for $ V_0= P^{e^{2r} \hat{X}^2 +  e^{-2r} \hat{Y}^2 \leq n_0+1}$, requiring $ n_0 \geq 9 e^{2|r|}\ln\left[\frac{12(k+n)}{k}\left(\frac{e^r}{1-q} +\frac{e^{-r}}{q}\right)\right]$. For a permutation invariant quantum state $ \rho^{2k+n}\in \mathit{S}(\mathcal{H}) $ with $ n>2k $, if $\{Z_1, Z_2,..., Z_k\}  $represent the measurement $ U^{\otimes k} $ on the k subsystems of $ \rho^{2k+n}$, and representing the event that the projection $ P^{k+n}_{\bar{\mathcal{H}}^{\otimes n}} $  on the left of $n+k$ subsystems fail as $ \mathit{F} $, then the probability of having more than $ \frac{n_0}{2} $ outcomes as $ U_1 $ while failing the projection is bounded by
\begin{eqnarray}
P[\left(\max_{i=1}^k Z_i^2 <\frac{n_0}{2} \right)\wedge \mathit{F}]\leq 8 k^{3/2}e^{-\frac{4q(1-q)k^3}{25(k+n)^2}}\label{Prob}
\end{eqnarray}

Proof:  Choose the deviation $ \delta=\frac{2q(1-q)k}{5(n+k)} $ in Lemma 1 and we can bound the gamma function plus deviation as 
\begin{widetext}
\begin{align}
\gamma_{U_1\to V_1}(\delta) + \delta & < \frac{2+q(1-q)}{q(1-q)}\delta +\frac{6}{q(1-q)}[q\cdot e^r\exp(-\frac{n_0 e^{-2r}}{9})+(1-q)\cdot e^{-r}\exp(-\frac{n_0e^{2r}}{9})]\\\label{Gamma2}
 & < \frac{k}{2(n+k)} + \left[ \frac{6}{(1-q)}e^r +\frac{6}{q}e^{-r}\right] e^{-n_0 e^{-2|r|}/q}\\
 & \leq \frac{k}{n+k}.
\end{align}
\end{widetext}
The probability in Eq.~(\ref{Prob}) can be rewritten as 
\begin{align}
&P[\left(f_{Z_{1},..., Z_{k}}=0 \right) \wedge \left(f_{Z_{k+1},..., Z_{k+n}}> \frac{k}{n+k}\right)
]\\
&\geq P[  f_{Z_{k+1},..., Z_{k+n}}> \frac{k}{n+k}]\\
&\geq P[  f_{Z_{k+1},...,Z_{k+n}}>\gamma_{U_1\to V_1}(\delta) + \delta
\end{align}  
Consequently, Lemma. 3, reduce the error probability to the right hand side of Eq.~(\ref{Prob}). As remarked in Remark 4 of \cite{renner2008}, the rest of the de Finetti theorem proof are identical. We thus generalizes the de Finetti theorem to different types of measurements, providing specific requirement for the experimental verification step. 
\begin{figure}[H]
  \includegraphics[width=15.5cm]{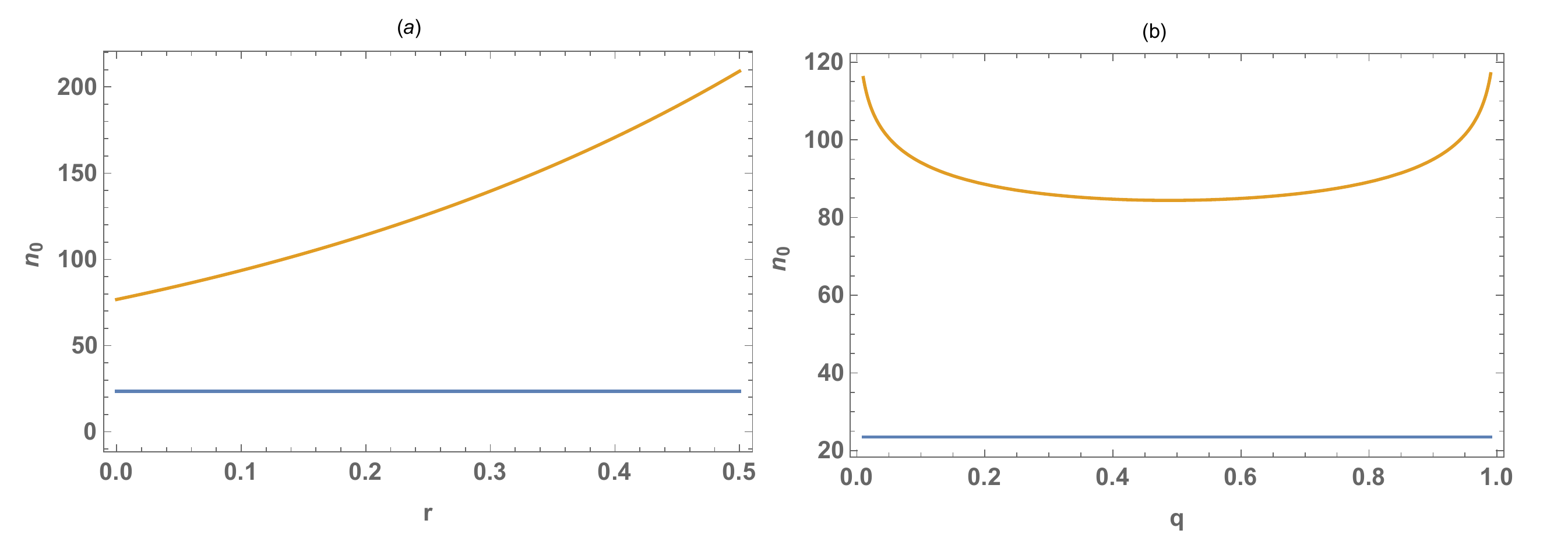}\\
  \caption{ In (a) yellow line represents the verification threshold $ n_0 $ dependence on squeezing strength $ r $ in the current work, and blue line represents the same relation in symmetric de Finetti theorem, both with  $k=2\times 10^7, n=2\times 10^9, q=0.4$. (b) shows the verification threshold $ n_0 $ dependence on basis selection probability $ q $ in the current work, and blue line represents the same relation in symmetric de Finetti theorem both with $x=0.05, k=2\times 10^7, n=2\times 10^9$.} \label{n0}
\end{figure}
The verification threshold $ n_0 $ which signifies the energy level of the state that could pass verification depends on both squeezing strength $ r $ and basis selection probability $ q $. We compare the $ n_0 $ value in this work with that derived in \cite{renner2008} without considering squeezing and biased basis selection in Fig.~\ref{n0}. We notice that in Fig.~\ref{n0} (a) our current energy bound is larger than that in \cite{renner2008} even at zero squeezing due to the strictly smaller bound in Eq.~(\ref{Gamma2}). In Fig.~\ref{n0} (b), the lowest energy shreshold is achieved at $ q=1/2 $ at non-zero squeezing which is still larger than the bound not considering squeezing. Our result thus validates the assumption used in \cite{renner2008} that experimental verification test designed for coherent state inputs lower bounds the energy threshold $ n_0 $ for general Gaussian state inputs.

\subsection{Discussions}\label{summary}

There are two new insights from our generalized de Finetti theorem. First, we introduce bases selection probability $ q $ to the experimental verification, demonstrating the extreme case when only one of the two conjugate bases is measured: at $ q=0 $  or $ q=1 $ the error probability is no longer  exponentially suppressed by $ k $ in Eq.~(\ref{Prob}). This result showcases the importance of keeping both conjugate measurements for verification to reduce unwanted correlation between subsystems and provide approximate i.i.d structure. Moreover, the error probability is minimized at $ q=1/2 $, which proves that the de Finetti theorem using symmetric basis selection in \cite{renner2008} gives a lower bound on error probabilities for general measurement strategies.   

Secondly, the squeezing parameter $ r $ gives a tighter bound on the quadrature variance threshold for verification given the input state is squeezed state. Although the success probability is independent of the squeezing strength of the states being prepared conditioned on passing the experimental verification on $ k $ subsystems, the value for the experimental verification threshold depends on $ r $  as $   n_0 \geq q e^{2|r|}\ln\left[\frac{12(k+n)}{k}\left(\frac{e^r}{1-q} +\frac{e^{-r}}{q}\right)\right]$, which increases monotonically with squeezing strength.  This implies that Alice and Bob should adjust their experimental verification energy threshold according to the states being transmitted through the quantum channel  using de Finetti theorem against coherent attack.

\bibliographystyle{apsrev4-1}
\bibliography{fano}

\end{document}